\documentclass[aps,prd,twocolumn,floatfix]{revtex4} 
\usepackage{epsfig}

\usepackage{epsfig}      

\def\g00{{\cal R}}
\def\grr{{\cal V}}
\def\Rs{{\cal R}_{\rm s}}
\def\Vs{{\cal V}_{\rm s}}
\def\zs{z_{\rm s}}

\begin{document}

\title{\textsc{Testing General Metric Theories of Gravity with
Bursting Neutron Stars}}

\author{Dimitrios Psaltis} \affiliation{ Departments of
Physics and Astronomy, University of Arizona, Tucson, AZ 85721}

\begin{abstract}
I show that several observable properties of bursting neutron stars in
metric theories of gravity can be calculated using only conservation
laws, Killing symmetries, and the Einstein equivalence principle,
without requiring the validity of the general relativistic field
equations. I calculate, in particular, the gravitational redshift of a
surface atomic line, the touchdown luminosity of a radius-expansion
burst, which is believed to be equal to the Eddington critical
luminosity, and the apparent surface area of a neutron star as
measured during the cooling tails of bursts.  I show that, for a
general metric theory of gravity, the apparent surface area of a
neutron star depends on the coordinate radius of the stellar surface
and on its gravitational redshift in the exact same way as in general
relativity. On the other hand, the Eddington critical luminosity
depends also on an additional parameter that measures the degree to
which the general relativistic field equations are satisfied. These
results can be used in conjunction with current and future high-energy
observations of bursting neutron stars to test general relativity in
the strong-field regime.
\end{abstract}

\date\today


\maketitle

\section{INTRODUCTION}

Black holes and neutron stars probe the strongest gravitational fields
found in the present-day universe. Astrophysical systems with
properties that are dominated by such compact objects are routinely
observed throughout the electromagnetic spectrum and will soon be
detected by gravitational wave observatories. It is our hope and
expectation that such observations will allow for clean tests of
strong-field general relativity and will demonstrate beyond doubt
that black-hole candidates are surrounded by event horizons.

Testing strong-field gravity and searching for evidence of event
horizons with current and future observations require a theoretical
framework within which such observations can be understood and
possible violations of the general relativistic predictions can be
quantified. Because of the strong gravitational field and the
relativistic velocities found in the vicinities of compact objects,
the Parametric Post-Newtonian framework~\cite{ppn}, which has been
very successful in performing weak-field tests, cannot be
used. 

Current studies are either based on phenomenological spacetimes within
the context of general relativity~\cite{multipole}, or employ theories
that are derived from parametric extensions of the Einstein-Hilbert
action~\cite{ST}. Albeit useful in performing null hypothesis
experiments and strong-field tests, these approaches do not
necessarily explore the whole range of possible violations of the
general relativistic predictions. 

In this paper, I show that several observable quantities of bursting
neutron stars can be calculated without requiring the validity of
Einstein's field equations. Instead, they can be derived from
conservation laws and Killing symmetries, assuming only the validity
of the Einstein equivalence principle.  This alleviates the need for a
general parametric theory with which these strong-field tests of gravity can
be performed. It also allows for a direct translation of observable
quantities such as redshifts, luminosities, temperatures, and
apparent surface areas, into values of the metric elements on the
surfaces of neutron stars (see \S\ref{sec:ns}).

In the calculations reported here, I assume only the validity of the
Einstein equivalence principle. This is one of the fundamental
building blocks of gravity theories and encompasses the concepts of
the equivalence between inertial and gravitational mass, of local
Lorentz invariance, and of local position invariance~\cite{ppn}. A
violation of the equivalence principle has severe implications for
many non-gravitational experiments. This has allowed solar-system and
laboratory experiments to place constraints on the amplitudes of such
violations that reach as low as one part in $10^{15}$~\cite{ppn},
justifying largely the assumption that the Einstein equivalence
principle is valid even in the strong gravitational fields found in
the vicinities of compact objects. Note here that the validity of the
Einstein equivalence principle does not imply the validity also of the
strong equivalence principle and hence, in a general metric theory, a
self gravitating body would not be following the same geodesics as the
test particles considered in this study.

Finally, I assume for simplicity that the neutron stars are not rapidly
spinning. This allows for the result to be expressed in terms of only
three parameters of the metric of each compact object. The calculation
can be extended in principle to incorporate effects of rapid
rotation. In that case, however, the results will depend on at least
two additional metric elements.

In \S\ref{sec:general}, I describe the formalism used in the
calculation and justify the relevant assumptions. In \S\ref{sec:ns}, I
calculate three observables related to the spectra of bursting neutron
stars that have been used in the past both in measuring their masses
and radii~\cite{nsMR} and in constraining scalar-tensor gravity
theories~\cite{ST}. They are: the gravitational redshift of atomic
lines from their surfaces, the touchdown luminosities of
radius-expansion bursts, and their apparent surface areas during the
cooling of the thermonuclear bursts.  I conclude in \S\ref{sec:concl}
with a discussion of the prospect of using the calculations reported
here in performing strong-field tests of general relativity with X-ray
observations of neutron stars.

\section{Assumptions and Definitions}
\label{sec:general}

In the following sections, I will calculate a number of observable
properties of a bursting neutron star, assuming only the validity of
the Einstein equivalence principle~\cite{ppn}, without relying on any
field equations for the underlying gravity theory. I will assume for
simplicity that the compact object is slowly spinning so that its
spacetime can be cast in the general form
\begin{equation}
ds^2=-\g00^2 dt^2+\grr^2 dr^2 + r^2 d\Omega^2\;.
\label{eq:metric}
\end{equation}
I will also assume that the redshift factor, $\g00$, and the volume
correction factor, $\grr$, are only functions of the coordinate radius
$r$ with the proper Minkowskian asymptotic limits.  In this spacetime,
I will denote the coordinate radius of the stellar surface by $r_{\rm
s}$ and the corresponding values of the metric elements there by $\Rs$
and $\Vs$. In general relativity, as well as in a number of
alternative gravity theories, $\g00\grr=1$ in the spacetime external
to the star and $\Rs\Vs \simeq 1$ in its low-density surface
layers. 

Because of the assumption of the validity of the equivalence
principle, matter and photons follow geodesics of the spacetime
described by the metric~(\ref{eq:metric}). I will describe the motion
of a massive particle in terms of its 4-velocity
$u^\mu\equiv(u^t,u^r,u^\theta,u^\phi)$. I will calculate the $u^t$
and $u^r$ components of the 4-velocity using the conservation laws
that arise from the two Killing vectors with components
$\xi^\mu=(1,0,0,0)$ and $\eta^\mu=(0,0,0,1)$ of the spacetime, i.e.,
the conservation of energy $E=-\xi^\mu u_\mu=\g00^2 u^t$ and of
angular momentum $l=\eta^\mu u_\mu=r^2 u^\phi$. Without loss of
generality, I will rotate the coordinate system so that the orbit of
the particle is at the equatorial plane, i.e., I will set
$\sin\theta=1$ and $u^\theta=0$. Finally, I will calculate the last
component of the 4-velocity of the particle using the requirement
$u_\mu u^\mu=-1$, which leads to
\begin{equation}
u^r=\left[\frac{E^2}{\g00^2\grr^2}-\frac{1}{\grr^2}
   \left(1+\frac{l^2}{r^2}\right)\right]^{1/2}\;.
\label{eq:vparticle}
\end{equation}

I will describe the motion of a photon in terms of its 4-momentum
$k^\mu=(k^t, k^r, k^\theta, k^\phi)$. Because of the same Killing
vectors and symmetries as in the case of massive particles, I can
again write $k^t=E/\g00^2$, $k^\theta=0$, and $k^\phi=l/r^2$. Finally,
I will calculate the radial component of the photon 4-momentum from
the requirement $k^\mu k_\mu=0$, which gives
\begin{equation}
k^r=\left(\frac{E^2}{\g00^2\grr^2}-\frac{1}{\grr^2}
   \frac{l^2}{r^2}\right)^{1/2}\;.
\label{eq:kphoton}
\end{equation}
Because the spacetime is asymptotically Minkowskian, by assumption,
the ratio $b\equiv l/E$ is the asymptotic impact parameter of the
photon at radial infinity.

In the following, I will also make frequent use of the 4-velocity of a
static observer at coordinate radius $r$, which is given by
\begin{equation}
u_{\rm obs}^\mu=\left(\frac{1}{\g00},0,0,0\right)\;.
\label{eq:obs}
\end{equation}

\section{Observed Properties of Neutron Stars}
\label{sec:ns}

Thermally emitting neutron stars are prime candidates both for
constraining the equation of state of neutron-star matter~\cite{nsMR}
and for testing general relativity in the strong-field
regime~\cite{ST}. This is especially true for neutron stars that
experience strong thermonuclear flashes on their surface layers, which
are observed as type~I X-ray bursts~\cite{burst}. 

The three observable properties of bursting neutron stars that can be
used in testing strong-field general relativity are the gravitational
redshift of a surface atomic line, the touchdown luminosity of a
radius-expansion burst, and the apparent surface area during the
cooling phases of the bursts. I will now calculate them in detail.

\subsection{Gravitationally Redshifted Lines}

If the spectrum of a thermally emitting neutron star has absorption
(or emission) features characteristic of atomic transitions, these
features will be detected by an observer at infinity with a
gravitational redshift equal to
\begin{equation}
   \zs\equiv\frac{\delta \lambda}{\lambda_0}=\Rs^{-1}-1\;,
   \label{eq:redshift}
\end{equation}
where $\lambda_0$ is the rest-frame wavelength of the atomic transition.
For a slowly-spinning neutron star, the emission line will be rotationally
broadened to a width $\Delta \lambda$ of
\begin{equation}
\Delta \lambda\simeq 2\frac{\Omega r_{\rm s}}{c}\;,
\end{equation}
where $\Omega$ is the spin frequency of the star.

\subsection{Touchdown Luminosity of a Radius-Expansion Burst}

The brightest among the type~I X-ray bursts from an accreting neutron
star show strong spectroscopic evidence for rapid expansion of the
radius of the X-ray photosphere~\cite{PRE}. It is widely believed that
the luminosities of these bursts reach the Eddington critical
luminosity at which the outward radiation force balances gravity,
causing the expansion of the surface layers of the neutron star. The
touchdown luminosities of radius-expansion bursts from a given source
remain constant between bursts to within a few percent, giving
empirical verification to the theoretical expectation that the emerging
luminosity is approximately equal to the Eddington critical
luminosity~\cite{PRE}. 

In general relativity, the Eddington critical luminosity at the
surface of bursting neutron star depends on its mass and radius. I
calculate here this critical luminosity for a compact object with an
external spacetime described by the metric~(\ref{eq:metric}), following
refs.~\cite{AEL90,LM95}.

I define the Eddington limit as the critical flux at which the
outward radiation force keeps the radial velocity of a particle
constant, i.e., the one for which
\begin{equation}
\frac{d^2r}{d\tau^2}=0\;,
\end{equation}
where $\tau$ is the proper time for the particle. If $f^\mu$ is the 
4-force that correspond to the Eddington limit, then from the geodesic
equation in the radial direction
\begin{equation}
\frac{1}{m}f^r=\frac{d^2r}{d\tau^2}+\Gamma^r_{\mu\nu}u^\mu u^\nu
     =\Gamma^r_{\mu\nu}u^\mu u^\nu\;,
\label{eq:geo}
\end{equation}
where $m$ is the rest mass of the particle and $u^\mu$ its 4-velocity.
The only non-zero component of the connection that enters the geodesic
equation~(\ref{eq:geo}) is
\begin{equation}
\Gamma_{tt}^r=-\frac{\g00}{\grr^2}\frac{\partial \g00}{\partial r}\;.
\end{equation}
For a particle at rest with a 4-velocity $u^\mu=(\g00^{-1},0,0,0)$, 
the radiation force that corresponds to the Eddington limit becomes
\begin{equation}
\frac{1}{m}f^r=-\frac{1}{\g00\grr^2}\frac{\partial \g00}{\partial r}\;.
\label{eq:acc}
\end{equation}

The outward radiation force is related to the radiation flux in the
particle's rest frame $F^r$ via
\begin{equation}
f^r=\sigma F^r\;,
\end{equation}
where $\sigma$ is the cross section for interaction. For a particle
at rest, the radiation flux is given by
\begin{equation}
F^r=-T^{(r)(t)}u_t=\g00 T^{(r)(t)}\;,
\label{eq:flux}
\end{equation}
where $T^{(r)(t)}$ is one element of the energy tensor of the
radiation field and the parentheses in the indices emphasize the fact
that no summation over the repeated index is implied. The $rt$-element
of the energy tensor of the radiation field at any coordinate radius
$r$ can be calculated from (ref.~\cite{AEL90} eq.~(3.32))
\begin{equation}
T^{(\hat{t})(\hat{r})}=\pi I(r) \sin^2\alpha\;,
\label{eq:Trt}
\end{equation}
where the hats on the indices indicate the fact that these are
the tetrad components of the tensor. Using the notation of Ref.~\cite{AEL90},
\begin{equation}
T^{(t)(r)}=\frac{1}{{\cal R}{\cal V}}T^{(\hat{t})(\hat{r})}=
\frac{\pi I(r) \sin^2\alpha}{{\cal R}{\cal V}}\;.
\end{equation}
Here $\alpha$ is the opening angle of the star as viewed by a static
observer at coordinate radius r, and $I(r_{\rm s})$ is the
frequency-integrated specific intensity of the radiation, which I
consider here to be isotropic within the opening angle $\alpha$.

\begin{figure}[t]
\epsfig{file=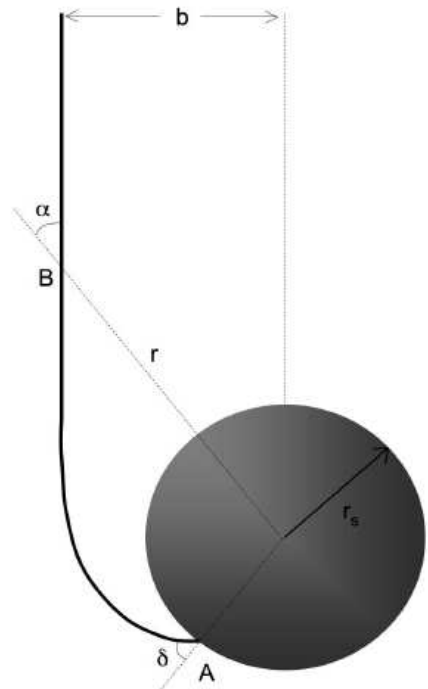, width=0.8\columnwidth}
\caption{\footnotesize
The geometry used in deriving the Eddington critical luminosity
for a slowly spinning neutron star.}
\label{fig:photons}
\end{figure} 

In order to calculate the opening angle $\alpha$, I will follow the
procedure of ref.~\cite{PFC83}, as illustrated in Figure~1.  Let point
B be the position of a static observer in the spacetime with a
4-velocity $u^\mu$ (eq.~(\ref{eq:obs})), and also of two photons with
4-momenta $w^\mu$ and $v^\mu$. The angle $\xi$ between the two photons
as measured by the static observer is
\begin{equation}
\cos\xi=1+\frac{v^\mu w^\mu}{(u^\nu w_\nu)(u^\sigma v_\sigma)}\;.
\end{equation}
I will consider one photon to be traveling radially outwards, i.e.,
with a 4-momentum vector that corresponds to $l=0$
(eq.~\ref{eq:kphoton}))
\begin{equation}
w^\mu=\left(\frac{\grr}{\g00},1,0,0\right)
\end{equation}
and the other to be is traveling on a non-radial null geodesic
\begin{equation}
v^\mu=E\left(\frac{1}{\g00^2},\left[\frac{1}{\g00^2\grr^2}-
    \frac{b^2}{\grr^2r^2}\right]^{1/2},0,\frac{b}{r^2}\right)\;,
\end{equation}
where $b\equiv l/E$. The angle between the two photons as measured
by the static observer is
\begin{equation}
\cos\xi=\left(1-\frac{\g00^2}{r^2}b^2\right)^{1/2}\;,
\end{equation}
from which it follows that
\begin{equation}
\frac{r}{\g00} \sin\xi =b
\end{equation}
is a constant of motion for each photon. This allows me to calculate the
opening angle of the star as viewed by an observer at coordinate radius
$r$. 

A photon that leaves the stellar surface at point A at an angle
$\delta$ with respect to the radial direction is detected by the
observer at radius $r$ traveling at an angle $\alpha$ with respect to
the local radial direction. In its motion, the photon follows a null
geodesic characterized by a single impact parameter $b$ and hence
\begin{equation}
b=\frac{r_s}{\Rs}\sin\delta=\frac{r}{\g00}\sin\alpha\;.
\end{equation}
If the stellar surface is at a coordinate radius larger than the one of
the photon orbit, then the photon with the largest angle $\alpha$
detected by the observer is the one that emerged from the stellar
surface at $\delta=\pi/2$. If the stellar surface is inside the photon
orbit then the opening angle corresponds to $\delta<\pi/2$.  As a
result, the opening angle of the source as seen by the static observer
is
\begin{equation}
\sin\alpha\le\frac{\g00}{\Rs}\frac{r_{\rm s}}{r}\;,
\label{eq:opening}
\end{equation}
with the equal sign corresponding to stars larger than the photon
orbit radius. Because I will be looking for small deviations from
general relativistic neutron stars, all of which are larger than the
radius of the photon orbit, I will consider here only this case.  It
is interesting that, although the details of self-lensing depend on
the proper volume factor $\grr$, the opening angle of the source as
seen by an observer at infinity depends only on the redshift factor
$\g00$, as in general relativity.

The radiation force on a particle becomes
\begin{equation}
f^r=\frac{\sigma \g00 \pi I(r)}{{\cal R}{\cal V}} 
\left(\frac{\g00}{\Rs}\right)^2
   \left(\frac{r_{\rm s}}{r}\right)^2\;.
\end{equation}
In order for the particle to remain at a constant radius $r$, this
force has to be equal to (see eq.~(\ref{eq:acc}))
\begin{equation}
\frac{1}{{\cal R}{\cal V}}
\frac{\sigma}{m}\g00 \pi I(r) \left(\frac{\g00}{\Rs}\right)^2
   \left(\frac{r_{\rm s}}{r}\right)^2=-\frac{1}{\g00\grr^2}
   \frac{\partial \g00}{\partial r}\;.
\label{eq:force}
\end{equation}
Evaluating this expression at the stellar surface, I obtain
\begin{equation}
\frac{\sigma}{m}\Rs^2 \pi I(r_{\rm s})=-\frac{\g00}{\grr}
   \left.\frac{\partial \g00}{\partial r}\right\vert_{r_{\rm s}}\;.
\label{eq:balance}
\end{equation}

I now connect the specific intensity of the radiation field at the
stellar radius with that at infinity. Conservation of the photon
occupation number
\begin{equation}
n\equiv \frac{I(r_{\rm s})}{g_{\mu\nu}k^\mu u_{\rm obs}^\nu}
\end{equation}
leads to 
\begin{equation}
I(r\rightarrow\infty)=I(r_{\rm s})\Rs^4\;,
\end{equation}
from which I can also infer for the luminosity $L_{\rm s}=4\pi r_{\rm s}^2
\pi I(r_{\rm s})$ that
\begin{equation}
L(r\rightarrow\infty)=L(r_{\rm s})\Rs^2\;.
\end{equation}
As a result, condition~(\ref{eq:force}) becomes
\begin{equation}
L_{\rm E}^\infty=\frac{4\pi m r_{\rm s}^2}{\sigma}
   \left(-\frac{\Rs}{\Vs}\left.\frac{\partial \g00}{\partial r}
   \right\vert_{r_{\rm s}}\right)=\frac{4\pi m}{\sigma}
    \frac{r_{\rm s}^2g_{\rm eff}}{(1+\zs)^2}\;,
\label{eq:Ledd}
\end{equation}
where I have defined $L_{\rm E}^\infty$ as the Eddington critical
luminosity as measured by an observer at infinity and the effective
acceleration at the stellar surface by
\begin{equation}
g_{\rm eff}=\frac{1}{\Rs\Vs}\left.
   \frac{d\g00}{dr}\right\vert_{r_{\rm s}}\;.
\end{equation}

\subsection{Apparent Surface Area During Burst Cooling}

A typical type~I X-ray burst on a neutron star is characterized by a
rapid ($\simeq 1$~s) rise and a slower (tens of seconds) decay, during
which the thermonuclear flash has spread throughout the entire neutron
star. Observations of the cooling tails of multiple type~I bursts from
a single source have shown that the apparent surface area of the
emitting region, defined as
\begin{equation}
S^\infty\equiv \frac{4\pi D^2 F_{\rm c,\infty}}{\sigma_{\rm SB}T_{\rm
c,\infty}^4}
\label{eq:S}
\end{equation}
remains approximately constant during each burst and between bursts
from the same source. In definition~(\ref{eq:S}), $F_{\rm c,\infty}$
is the measured flux of the source during the cooling tail of the
burst, $T_{\rm c}^\infty$ is the measured color temperature of the
burst spectrum, $D$ is the distance to the source and $\sigma_{\rm
SB}$ is the Stefan-Boltzmann constant.  In general relativity, the
apparent surface area during the cooling tails of X-ray bursts from a
neutron star depends on the stellar mass and radius. I calculate here
this same quantity for an object described by a general spherically
symmetric metric.

The color temperature on the surface of the neutron star is related to
the color temperature measured at infinity by a simple redshift, i.e.,
\begin{equation}
T_{\rm c,s}=\frac{T_{\rm c,\infty}}{\Rs}\;.
\end{equation}
I can then use equations~(\ref{eq:flux}), (\ref{eq:Trt}), and
(\ref{eq:opening}) to calculate the radiation flux measured at
infinity as
\begin{equation}
F_{\rm c,\infty}=\pi I(r_{\rm s})\Rs^2 
   \left(\frac{r_{\rm s}}{D}\right)^2\;.
\end{equation}
If the spectrum of a neutron star during the cooling tail of a burst
were a pure blackbody, then $\pi I(r_{\rm s})=\sigma_{\rm SB} [T(r_{\rm
s})]^4$. However, the interaction of photons with the atmospheric
electrons and atoms distort the shape of the emerging spectrum. The
distortion is typically parameterized in terms of a color correction
factor
\begin{equation}
f_{\rm c}\equiv \frac{T_{\rm c}}{T_{\rm eff}}\;,
\end{equation}
where $T_{\rm c}$ and $T_{\rm eff}$ are the color temperature and the
effective temperatures of the atmosphere, respectively. The value of
the color correction factor can be calculated a priori and depends on
the temperature and the composition of the atmosphere as well as on
the local acceleration~\cite{color}. With this definition,
\begin{equation}
\pi I(r_{\rm s})=\frac{\sigma_{\rm SB}T_{\rm c,s}^4}{f_{\rm c}^4}
\end{equation}
and the observed ratio $F_{\rm c,\infty}/\sigma_{\rm SB} T_{\rm c,\infty}^4$
becomes
\begin{equation}
\frac{F_{\rm c,\infty}}{\sigma_{\rm SB} T_{\rm c,\infty}^4}
= \frac{1}{f_{\rm c}^4 \Rs^2} \frac{r_{\rm s}^2}{D^2}=
    \frac{1}{f_{\rm c}^4}
    \left(\frac{r_{\rm s}}{D}\right)^2
    (z_{\rm s}+1)^2\;.
\end{equation}
It is worth emphasizing that the dependence of the observable
on the coordinate radius of the stellar surface and on the redshift
is the same as in the case of general relativity.

\section{Testing General Relativity with Bursting Neutron Stars}
\label{sec:concl}

In \S\ref{sec:ns}, I showed that various observables from a slowly
spinning neutron star depend on three parameters related to the
coordinate radius of its surface and the values of the metric elements
there. In general relativity, however, all these observables depend on
only two parameters: the gravitational mass of the compact object and
the coordinate radius of its stellar surface. The difference in the
number of free parameters between a general metric theory and general
relativity makes a direct test of the latter theory possible.  Indeed,
the three parameters of a compact object in a general metric theory can
be measured (or at least constrained) using high-energy
observations. If general relativity describes accurately gravitational
phenomena in the strong-field regime, then the three parameters cannot
be independent but they must satisfy a consistency relation, which
I calculate below. The degree to which consistency can be demonstrated
will provide a measure of the degree to which general relativistic
predictions have been tested.

In general relativity, the two components of the
metric~(\ref{eq:metric}) at the stellar surface are related to the
mass $M$ and the radius $r_{\rm s}$ of the compact object by
\begin{equation}
\Rs=\Vs^{-1}=\left(1-\frac{2M}{r_{\rm s}}\right)\;.
\end{equation}
Instead of using the mass of the compact object as the second
independent variable, I will choose the redshift from its surface,
\begin{equation}
z_{\rm s, GR}=\left(1-\frac{2M}{r_{\rm s}}\right)^{-1}-1\;.
\end{equation}
The effective gravitational acceleration in the surface layers of 
a general relativistic star then becomes
\begin{equation}
g_{\rm eff, GR}=\frac{1}{2r_{\rm s}}
\left[\frac{z_{\rm s}(z_{\rm s}+2)}{z_{\rm s}+1}\right]\;.
\end{equation}

I can now define a parameter $\eta$ to measure
deviations from general relativity by
\begin{equation}
g_{\rm eff}=\eta g_{\rm eff, GR}= \eta
\frac{1}{2r_{\rm s}}
\left[\frac{z_{\rm s}(z_{\rm s}+2)}{z_{\rm s}+1}\right]\;.
\end{equation}
This parameter quantifies the degree to which the gravitational
acceleration on the surface of a star can be determined entirely
by its mass and radius, as predicted by general relativity.

With the above definitions, I can write the three observable
quantities for a bursting neutron star of known distance, $D$, as
\begin{eqnarray}
\frac{\delta \lambda}{\lambda_0}&=&z_{\rm s}\label{eq:obs1}\\
L_{\rm E}^\infty&\equiv&
   4\pi D^2 F_{\rm E}^\infty=\frac{4\pi m_{\rm p} r_{\rm s}}
   {(1+X)\sigma_{\rm T}} 
   \left[\frac{z_{\rm s}(z_{\rm s}+2)}{(1+\zs)^3}\right] 
   \eta\label{eq:obs2}\\
R_{\rm app}&\equiv&
D\left(\frac{F_{\rm c,\infty}}{\sigma_{\rm SB} T_{\rm c,\infty}^4}
  \right)^{1/2}=
    \frac{r_{\rm s}}{f_{\rm c}^2}
    (z_{\rm s}+1)\label{eq:obs3}\;,
\end{eqnarray}
where I have introduced the apparent source radius $R_{\rm app}$. I
have also assumed that the interaction between the electrons and the
photons is due to coherent Thomson scattering and denoted the hydrogen
mass fraction of the accreted material by $X$. 

It is important to note that, for a general metric theory of gravity,
the first and the third observables depend on the coordinate radius of
the neutron-star surface, $r_{\rm s}$, and on its gravitational
redshift, $z_{\rm s}$, in the exact same way as in general
relativity. On the other hand, the Eddington luminosity depends also
on the parameter $\eta$, which measures possible deviations from
general relativity. Therefore, it is only through measurements of the
touchdown luminosity of radius expansion bursts that general
relativity can be tested with bursting neutron stars. 

In general relativity, $\eta_{\rm GR}=1$, by definition. On the other
hand, in other gravity theories, this parameter depends also on the
additional gravitational degrees of freedom. For example, in a
scalar-tensor gravity as the one described in ref.~\cite{ST}), $\eta$
depends on the magnitude of the scalar field at the surface of the
neutron star, which in turn depends on the mass distribution inside
the star and its coupling to the scalar field.

Not all combinations of values of the three observables are possible
in general relativity.  Buchdahl's theorem states that no spherical,
general relativistic star can exist with a mass-to-radius ratio larger
than $M/R>4/9$. This corresponds to a maximum possible value of the
redshift from the surface of a general relativistic star of $z_{\rm
s}\le 2$. The apparent radius during the cooling tails of bursts
depends monotonically on the redshift and, therefore, reaches the
highest possible value in general relativity at $z_{\rm
s}=2$. Finally, the Eddington luminosity is a non-monotonic
function of redshift and peaks at $z_{\rm s}=\sqrt{3}-1\simeq
0.73$. The maximum possible values of the three observables for a
general relativistic star are, therefore,
\begin{eqnarray}
\left.\frac{\delta \lambda}{\lambda_0}\right\vert_{\rm GR}&\le&2\\
L_{\rm E,GR}^\infty&\le&
   (\sqrt{3}-1)\frac{4\pi m r_{\rm s}}{(1+X)\sigma_{\rm T}} \\
R_{\rm app,GR}
    &\le&\frac{3}{f_{\rm c}^2}r_{\rm s}\;.
\end{eqnarray}
Violation of any of the above inequalities will also signify
that general relativity does not describe accurately gravitational
phenomena in the strong-field regime.

\begin{figure}[t]
\epsfig{file=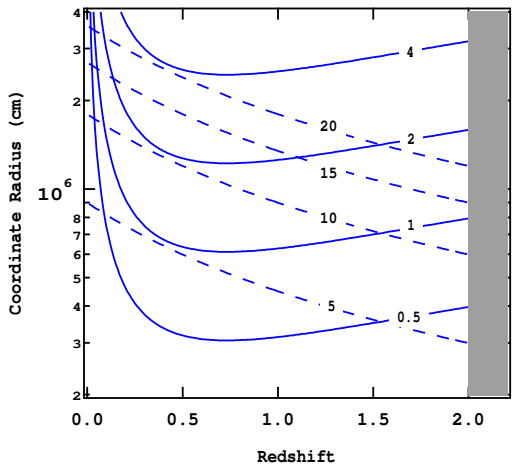, width=0.9\columnwidth}
\caption{\footnotesize {\em Solid lines:\/} Contours of constant
Eddington luminosity at infinity, in units of $10^{38}$~erg~s$^{-1}$, as
a function of the coordinate radius and the redshift of the surface of
the bursting neutron star {\em Dashed lines:\/} Contours of constant
apparent radius of the neutron star, in units of 1~km, as inferred from
the cooling tail of a burst, on the same parameter space. In this calculation
the atmosphere of the neutron star was assumed to consist of pure hydrogen
and the color correction factor was set to 1.34.}
\label{fig:contours}
\end{figure}

A second test of general relativity can be obtained using a
slowly-spinning, bursting neutron star at a known distance that
accretes matter of known composition and shows radius-expansion type~I
X-ray bursts. Assuming the validity of general relativity,
equations~(\ref{eq:obs2}) and (\ref{eq:obs3}) can be inverted to yield
the coordinate radius of the neutron-star surface and its gravitational
redshift (see Fig.~\ref{fig:contours}). However, because the Eddington
luminosity does not have a monotonic dependence on redshift, this inversion
procedure is not always possible. For example, Figure~\ref{fig:contours}
shows that an apparent radius of 5~km and an Eddington luminosity of
$2\times 10^{38}$~erg~s$^{-1}$ are impossible in general relativity.

In order to calculate the range of simultaneously determined values
of the Eddington luminosity and of the apparent radius that are consistent
with general relativity, I solve equation~(\ref{eq:obs2}) for $r_s$ and
substitute it into equation~(\ref{eq:obs3}). After a small
rearrangement of terms, the result is
\begin{equation}
\frac{(1+X)\sigma_{\rm T}L_{\rm E,GR}^\infty}{4\pi m_{\rm p} f_{\rm c}^2
R_{\rm app,GR}}=\frac{z_{\rm s}(z_{\rm s}+2)}{(z_{\rm s}+1)^4}\;.
\end{equation}
The right-hand-side of this last equation has a maximum value of $1/4$ at
$z_{\rm s}=\sqrt{2}-1$, so that
\begin{equation}
L_{\rm E,GR}^{\infty}\le \frac{\pi m_{\rm p}f_{\rm c}^2}{(1+X)\sigma_{\rm T}}
R_{\rm app,GR}\;,
\end{equation}
which I can rewrite in CGS units as
\begin{equation}
L_{\rm E,GR}^{\infty}\le 1.9\times 10^{38}
   \left(\frac{f_{\rm c}}{1.34}\right)^2
   \left(\frac{2}{1+X}\right)
   \left(\frac{R_{\rm app,GR}}{10^6~\mbox{cm}}\right)
   ~\mbox{erg~s}^{-1}\;.
\label{eq:limits}
\end{equation}
Violation of inequality~(\ref{eq:limits}) will also be evidence for
new gravitational physics, not described by general relativity.

The optimal test of general relativity will occur from the detection
of gravitationally redshifted lines from a neutron star that is slowly
spinning, lies at a known distance, and shows radius-expansion type~I
X-ray bursts. In this case, equations~(\ref{eq:obs1})-(\ref{eq:obs3})
can be combined to give
\begin{equation}
\eta=\frac{(1+X)\sigma}{m f_{\rm c}^2}
F_{\rm E}^\infty 
\left(\frac{F_{\rm c,\infty}}{\sigma T_{\rm c,\infty}^4}\right)^{-1/2}
\frac{(1+\delta \lambda/\lambda_0)^4}{(2+\delta \lambda/\lambda_0)
(\delta \lambda/\lambda_0)}D\;.
\end{equation}
All the quantities on the right-hand side of this last equation are
either observable or can be calculated from models of neutron-star
atmospheres. The parameter $\eta$ is a measure of the
degree to which general relativity accurately describes the strong
gravitational fields found in the vicinities of neutron stars.

A successful application of the line of arguments discussed here
relies on the detection of radius-expansion bursts and of
gravitationally redshifted lines from sources with known distances.
Radius expansions bursts have been detected to date from at least 31
bursting neutron stars~\cite{data}, several of which lie in globular
clusters and, therefore, have also well constrained
distances~\cite{PRE}. The touchdown luminosities and apparent radii
can be inferred from current data to an accuracy of $\simeq 5\%$ that
is limited by systematic uncertainties only~\cite{PRE}. The distances
to the globular clusters that contain known bursters are known to an
accuracy of $\simeq 5-20\%$. Assuming the latter accuracy to be 10\%,
and assuming also that the composition of the accreting material can be
estimated to the same accuracy by long-wavelength observations of the
accretion flow, the proposed test can lead to a measurement of the
parameter $\eta$ to within $\simeq 15$\%.

Note, however, that gravitationally redshifted atomic lines have been
reported only for one, slowly rotating, burster,
EXO~0748$-$676. Future observations of bursting neutron stars with
telescopes that combine a high spectral resolution and a large
collecting area, such as Constellation-X and XEUS, will allow for
quantitative tests of general relativity in the strong-field regime
with bursting neutron stars.

\bibliographystyle{apsrev}

\end{document}